%% file: main.tex
\documentclass[10pt,conference]{IEEEtran}
\IEEEoverridecommandlockouts
\usepackage{graphicx,epstopdf}
\usepackage{url}
\usepackage{amsmath,amssymb}
\usepackage{graphicx}
\usepackage{textcomp}
\usepackage{color}
\usepackage{xcolor}
\usepackage{hyperref}
\usepackage{makecell}
\hypersetup{
	colorlinks   = true,
	urlcolor     = blue, %Colour for external hyperlinks
	linkcolor    = blue, %Colour of internal links
	citecolor   = blue %Colour of citations
}
\usepackage{listings}
\usepackage[numbers]{natbib}
\usepackage{multirow}
\usepackage{algpseudocode}
\usepackage[vlined,ruled,linesnumbered]{algorithm2e}
\usepackage{xspace}
\usepackage{enumitem}
\usepackage[misc]{ifsym}
\usepackage{stmaryrd}
\usepackage{tcolorbox}
\usepackage{nicematrix}
\usepackage{tikz}
\newcommand*\circled[1]{\tikz[baseline=(char.base)]{
		\node[shape=circle,fill=black,text=white,draw,inner sep=.15pt] (char) {#1};}}
		
\newcommand{\tool}{\textit{WebExplor}\xspace}
\DeclareMathOperator*{\argmax}{arg\,max}

\newtheorem{definition}{Definition}
\def\BibTeX{{\rm B\kern-.05em{\sc i\kern-.025em b}\kern-.08em
    T\kern-.1667em\lower.7ex\hbox{E}\kern-.125emX}}

\begin{document}
\title{Automatic Web Testing Using Curiosity-Driven Reinforcement Learning\\
\thanks{\Letter~Corresponding author: Xiaofei Xie.}
\thanks{ $^{*}$ Yan Zheng and Yi Liu contributed equally to this work.}
}

\author{
    Yan Zheng$^{\dagger,\ddagger,*}$, Yi Liu$^{\ddagger,\Lbag,*}$, Xiaofei Xie$^{\ddagger,\text{ \Letter}}$, Yepang Liu$^{\Lbag}$, Lei Ma$^{\divideontimes}$, Jianye Hao$^{\dagger}$, and Yang Liu$^{\ddagger}$\\
    
    $^\dagger$Tianjin University, Tianjin, China;
    $^\ddagger$Nanyang Technological University, Singapore\\
    $^\Lbag$Dept. of Comp. Sci. and Engr., Guangdong Provincial Key Laboratory of Brain-inspired Intelligent Computation,\\ Southern University of Science and Technology, Shenzhen, China;
    $^\divideontimes$Kyushu University, Fukuoka, Japan.\\
    % $^\text{\textsection}$Zhejiang Sci-Tech University, Hangzhou, China.\\
    
    % yanzheng@tju.edu.cn, 11610522@mail.sustech.edu.cn, \{xiaofei.xie,yangliu\}@ntu.edu.sg, 
    % \\liuyp1@sustech.edu.cn, 
    % malei@ait.kyushu-u.ac.jp, jianye.hao@tju.edu.cn
}

\maketitle
\begin{abstract}
	Web testing has long been recognized as a notoriously difficult task.
	Even nowadays, web testing still heavily relies on manual efforts while automated web testing is far from achieving human-level performance. Key challenges in web testing include dynamic content update and deep bugs hiding under complicated user interactions and specific input values, which can only be triggered by certain action sequences in the huge search space.
	In this paper, we propose \tool, an automatic end-to-end web testing framework, to achieve an adaptive exploration of web applications.
	\tool adopts curiosity-driven reinforcement learning to generate high-quality action sequences (test cases) satisfying temporal logical relations.
	Besides, \tool incrementally builds an automaton during the online testing process, which provides high-level guidance to further improve the testing efficiency.
	We have conducted comprehensive evaluations of \tool on six real-world projects, a commercial SaaS web application, and performed an in-the-wild study of the top 50 web applications in the world. The results demonstrate that in most cases \tool can achieve significantly higher failure detection rate, code coverage and efficiency than existing state-of-the-art web testing techniques. \tool also detected 12 previously unknown failures in the commercial web application, which have been confirmed and fixed by the developers.
	Furthermore, our in-the-wild study further uncovered 3,466 exceptions and errors.
% 	are discovered during the in-the-wild study.

% 	or search-based solutions. , automated web testing is still unable to achieve human-level performance in generating test cases. Key challenges include dynamic content update and deep bugs hiding after a certain action sequence.
% 	In this paper, we propose \tool, an automatic end-to-end web testing framework, which adopts curiosity-driven reinforcement learning to achieve an adaptive exploration of web applications and generate high-quality action sequences (test cases) with temporal logical relations.
% 	\tool incrementally builds an automaton during online testing, which provides high-level guidance to further boost the testing efficiency. \lei{more feature of the tool to highlight?}
% 	We have conducted a large-scale and comprehensive evaluation on six real-world projects, a commercial SaaS web application, and an in-the-wild study of the top 50 web applications in the world. The results demonstrate that \tool can mostly achieve significantly higher fault detection rate, better code coverage and efficiency than existing web testing techniques. \tool found 10 previously unknown faults in the commercial web application, which have been confirmed by the developers. Moreover, 3,290 exceptions and errors are discovered during the in-the-wild study.
\end{abstract}

% \begin{IEEEkeywords}
% Automatic web testing, reinforcement learning, curiosity, deterministic finite automaton.
% \end{IEEEkeywords}
\input{intro}
\input{bg}
\input{alg}

\input{evaluation}
\input{study}

\section{Conclusion}
This paper proposes \tool, a new RL based approach for automatic web testing. \tool leverages the curiosity-driven RL to achieve efficient and adaptive exploration. Meanwhile, DFA is proposed to provide high-level guidance, so as to further boost the testing efficacy.
% Experiments on benchmarks and real-word applications show that
% \tool achieves a better performance in terms of the exploration efficiency and failure detection.
Experiments show that \tool outperforms existing web testing techniques in both code coverage and failure detection.
%\tool achieves competitive performance compared with the state-of-the-arts
% strong bug detection
% more bugs are more likely to be discovered.
%\tool also demonstrates its usefulness on various real-world web applications.
In the future, we plan to extend \tool with hierarchical RL for a better exploration, and extract business scenarios from the DFA, whereby fault localization can be achievable.
\section{Acknowledgements}
The work is supported in part by the National Natural Science Foundation of China (Grant No. U1836214), Special Program of Artificial Intelligence and Special Program of Artificial Intelligence of Tianjin Municipal Science and Technology Commission (No. 569 17ZXRGGX00150), Tianjin Natural Science Fund (No. 19JCYBJC16300), Research on Data Platform Technology Based on Automotive Electronic Identification System, Singapore National Research Foundation, under its National Cybersecurity R\&D Program (No. NRF2018NCR-NCR005-0001), Singapore National Research Foundation under NCR (No. NRF2018NCR-NSOE003-0001), NRF Investigatorship (No. NRFI06-2020-0022), JSPS KAKENHI (Grant No. 20H04168, 19K24348, 19H04086), JST-Mirai Program (Grant No. JPMJMI18BB), Japan, and Guangdong Provincial Key Laboratory (Grant No. 2020B121201001), China.

\bibliographystyle{IEEEtran}
\bibliography{ref}
\end{document}

%% file: intro.tex
\section{Introduction}
The past decades have witnessed the unprecedentedly rapid development and innovation of web technologies.
% , transforming web applications into a completely new stage.
Nowadays, web applications have become as powerful as native desktop applications. They are competitively convenient and do not require complicated installation either. However, web applications can be difficult to test due to their complicated business logic implemented in different languages across the client and server side (e.g., HTML, JavaScript, C\# and Java).
In general, the more web pages are explored with more states covered, the higher the possibility of discovering defects becomes. Hence, various kinds of approaches have been proposed to achieve a sufficient exploration by generating test cases.

Manual designing with the aid of automation frameworks such as Selenium~\cite{huggins2018selenium} is a useful way to create test cases. %To verify the correctness of a web application，
The tester is required to create test scripts, simulating user operations (e.g., clicking buttons and filling in forms) on the web application’s graphical user interface (GUI)~\cite{biagiola2019web,stocco2018visual,leotta2018pesto,biagiola2017search,leotta2016robula+,hammoudi2016waterfall,leotta2015using,binder1996testing,fewster1999software}. However, such manual work is labor-intensive and costly, where the testing effectiveness heavily depends on the human testers' domain knowledge.
Besides, web applications frequently evolve and the manually written test cases normally require substantial modifications before testing the new versions~\cite{thummalapenta2013efficient,StoccoLRT17}.

\begin{figure}[t]
	\centering
	\includegraphics[width=0.8\linewidth]{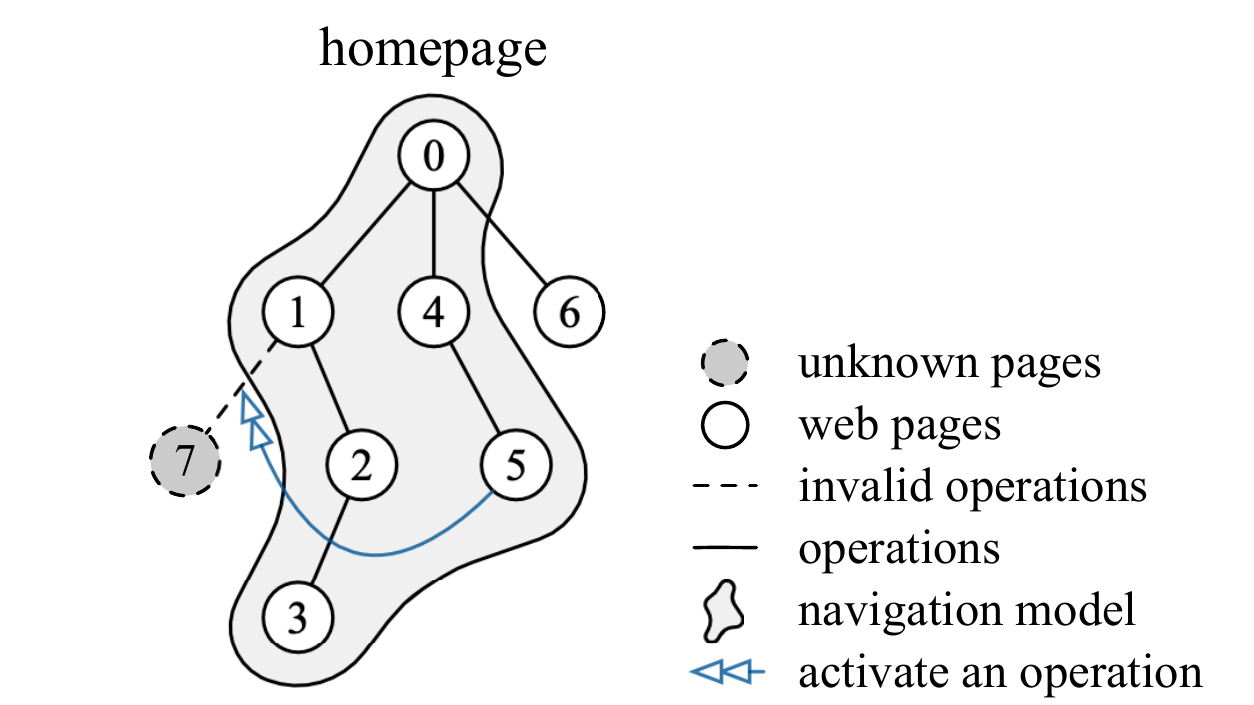}
	\caption{An intuitive visualization of exploring web pages.}
	\label{fig:motivation}
	\vspace{-5mm}
\end{figure}
Random-based approaches~\cite{monkey,crawljax} generate pseudo-random operations to fuzz the web applications. Despite the wide adoption in practical development, the shortcomings of such approaches are obvious. That is, they often create invalid test cases like performing input operations on buttons. Also, the testing is unbalanced and some hard-to-reach web pages may never be explored.

%To achieve automatic web testing, the state-of-the-art techniques mostly follow model-based web testing
Model-based approaches~\cite{FSE2019-WebTest,mesbah2012crawling,marchetto2008state,model-issta,model2-issta} build a navigation model of the web application under testing, and then generate test cases accordingly by random or sophisticated search strategies. In spite of the guidance of the navigation model, existing approaches still suffer from several limitations. 
Firstly, the constructed navigation model may cover only a part of the web application, restricting the exploration power of the generated test cases.
As depicted in Fig.~\ref{fig:motivation}, pages~6 and 7 cannot be tested as they are not included in the navigation model.
Also, domain knowledge is usually required in building high-quality models~\cite{FSE2019-WebTest}.
Besides, the content of web applications is usually dynamically updated (e.g., via JavaScript code execution), which cannot be easily captured by the static navigation models.
Secondly, in web applications, long sequences of actions (e.g., path $0 \rightarrow 1 \rightarrow 2 \rightarrow 3 $) are often needed to complete certain tasks such as filling in and submitting forms. The business logic of real-world web applications can be arbitrarily complex.
For example, page 7 can be visited only when page 5 is properly navigated. It could be challenging for random or search-based strategies~\cite{biagiola2017search,FSE2019-WebTest,ZhengWFCY18} to generate effective action sequences.

To address the aforementioned challenges, an effective and end-to-end automatic testing is needed.
Recently, reinforcement learning (RL) has demonstrated its potential for learning a policy to test and interact with complicated games~\cite{mnih2015human,ZhengMHZYF18,zheng2019wuji,ShenZHMCFL20} or Android applications~\cite{pan2020reinforcement,AdamoKKB18,VuongT18}, which provides the possibility of applying RL on automatic web testing.
However, existing techniques cannot be easily adapted to test web applications for the following reasons.
Firstly, the testing domains are different, which make the RL modeling totally different. For example, the reward function may be different. Game playing usually has 
concrete goals to achieve (e.g., winning the game or maximizing a score), which is not the case in web testing. The state definition and abstraction are different either. Game playing~\cite{zheng2019wuji} defines the state based on the outputs of APIs and Android testing~\cite{pan2020reinforcement} uses an existing tool UIAutomator~\cite{uiautomator} to extract structures as the states, which are both not applicable in web testing.
Secondly, one fundamental challenge of RL is how to perform effective exploration especially when the space of the environment is huge~\cite{dota,ZhengMHZ18}. The existing techniques~\cite{mnih2015human,zheng2019wuji,ZhengHZMH20} mainly guide the exploration with simple reward functions, which could be ineffective for web applications that have complex business logic and frequent dynamic update. Thus, more effective exploration is needed for RL-based web testing.

Considering the dynamic and highly interactive nature of web applications, an effective model-free web testing technology can be highly desirable.
In this paper, we propose a novel web testing framework, named \tool, which performs an end-to-end automated web testing.
\tool leverages RL to perform an adaptive exploration of web applications and generate high-quality action sequences, which may be prerequisite operations (e.g., filling forms before submission) for discovering new pages.
In particular, \tool performs an on-the-fly testing while constantly training the agent policies (rather than the usual AI solutions that can only be used after training).
To achieve both high coverage and efficiency during testing, we first propose the state abstraction based on the HTML pages. Then, we propose a curiosity-driven reward function, which provides low-level guidance for the exploration of RL such that the learned policy could explore more behaviors of the web applications. To avoid falling into local optima, especially when the learning space is huge, we further propose a deterministic finite automaton (DFA) guided exploration strategy that provides high-level guidance for RL to efficiently explore the web applications. In particular, the DFA records the transitions and states visited during the RL exploration and is continuously updated. When RL gets stuck (i.e., cannot find a new state within a given time budget), \tool selects one path from DFA based on the curiosity and guides RL to explore along this path further. Both the low-level guidance (from the reward function) and the high-level guidance (from the DFA) play important roles in achieving effective web testing.
To demonstrate the effectiveness of our technique, we implemented \tool and conducted a large-scale evaluation on a research benchmark of six real-world projects~\cite{FSE2019-WebTest} and a commercial SaaS web application. We also conducted an in-the-wild study of the top 50 web applications in the world~\cite{aleax}.
The contributions of this paper are summarized as follows.
\begin{itemize}
	\item We propose a novel web testing framework, \tool, to efficiently and effectively test real-world web applications. To the best of our knowledge, \tool is the first end-to-end web testing framework leveraging reinforcement learning.
	\item We propose a curiosity-driven reward function and a DFA to guide RL to efficiently explore diverse behaviors of web applications.
	\item We comprehensively evaluate \tool on six open-source web applications, a commercial web application, and top 50 real-world web applications. In the commercial application, $\underline{12}$ previously unknown failures, including logical and security defects, are discovered by \tool, and confirmed and fixed by the developers. Furthermore, $\underline{3,466}$ exceptions and errors are discovered in the top 50 web applications.
\end{itemize}

%% file: bg.tex
\section{Preliminaries}
\subsection{Web Application and Reinforcement Learning}
\begin{figure}[t]
	\centering
	\includegraphics[width=\linewidth]{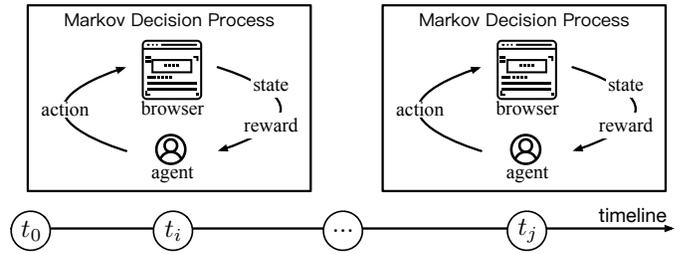}
	\caption{Web interactions as a Markov decision process.}
	\label{fig:mdp}
	\vspace{-4mm}
\end{figure}

A typical web application requires the end user to input a sequence of actions (e.g., clicks) to interact, which, in turn, will change the web application's states (e.g., URL or GUI). 
% The interaction continues unless the user stops. 
This process can be modeled as a Markov Decision Process (MDP)~\cite{sutton2018reinforcement}. MDP can be defined as a 4-tuple $ \left\langle \mathcal{S},\mathcal{A},\mathcal{R}, \mathcal{P} \right\rangle $, where $ \mathcal{S},\mathcal{A} $ represent the sets of states and actions, respectively.
As shown in Fig.~\ref{fig:mdp}, the agent (tester) interacts with the environment (browser) over the time horizon. At timestamp $t$,
%the state $ s_t \in \mathbf{S}$ is the observation of the environment  which is usually different at different timestamps.
the agent observes the state $ s_t \in \mathcal{S}$ of the web applications (e.g., GUI or HTML), and selects an action $ a_t \in \mathcal{A} $ to execute, after which, the agent receives an immediate reward $r_t  = \mathcal{R}(s_t, a_t)$, and the environment can change to a new state $ s_{t+1}\sim \mathcal{P}(s_t, a_t) $. The $ \mathcal{R}(\cdot) $ and $ \mathcal{P(\cdot)} $ are reward function and probability transition function, both of which depend only on preceding $ s_t $ and $ a_t $.

%The $ r_t $ and $ s_{t+1} $ have well-defined probability distributions dependent only on preceding state and action. That is, $ s_{t+1} \sim \mathcal{P}(s_t, a_t) $ and $ r_t = \mathcal{R}(s_t, a_t)$.

%The action $ a_t \in \mathbf{A} $ is what agent take (e.g., click, input) from a valid action sets at the moment. The reward $ r_t = \mathbf{R}(s_t, a_t)$ (e.g., scalar value) is the feedback measuring the quality of taking action $ a_t $ with regard to achieving some goals (e.g., filling forms or achieving URL jumping).

% depicts an MDP which shows a typical overall interaction of web applications. In general, the MDP consists of 5 elements - Agent, Environment, State, Action and Reward. An environment is what an agent (user) interacts with (e.g., a web application). The agent utilizes a specific policy for interacting with the environment over the time horizon. At timestamp $t$, the state is the observation of the environment (e.g., GUI/HTML) which is usually different at different timestamps. The action is what agent take (e.g., click, input) from a valid action sets at the moment. The reward (e.g., scalar value) is the feedback measuring the quality of an agent’s action with regard to achieving some goals (e.g., filling forms or achieving URL jumping).

Intelligently, the agent selects an action $a \sim \pi(s)$ to execute according to a probabilistic policy function $ \pi(\cdot) $. The agent interacting with the environment gives rise to a trajectory as follows:
\begin{equation}\label{eq:traj}
traj = (s_0, a_0, r_0, \cdots, s_t, a_t, r_t, \cdots),
\end{equation}
where the subscripts denote different timestamps over the finite time horizon. Each trajectory has a return, defined as $ \sum_{t=0}^{T} \gamma^t r_t$, where rewards are discounted by a factor $ \gamma \in [0,1)$. In general, RL aims at finding one optimal trajectory with the maximum return rather than diverse trajectories. However, this is often not the goal of web testing, which seeks to explore diverse trajectories and states. 
Therefore, an adaptive reward function is required to guide RL to continuously generate diverse trajectories, which will be detailed in Section~\ref{sec:alg}.

% That is trying to explore  via diverse trajectories.

%\begin{definition}[\textbf{Policy}]
%The agent uses a policy $ \pi(\cdot) $ that takes as into the state $s$ and outputs the action $a \sim \pi(s)$.
%\end{definition}
%Similar to the human interacting with the web application,
%Given different web states, policy $ \pi $ will select different actions (DOM elements) to execute according to certain probability. This sequential interaction creates a test case for web testing.

%To be specific, given the web state $s_t$ at time $t$ (e.g., GUI), the agent selects an action $a_t \sim \pi(s_t)$ to interact with the web browser environment and receives a reward $r_t$ from the environment. The environment turns into a new state $s_{t+1}$, affecting the agent selection of the next action $a_{t+1}$. The agent and environment interact continually until the agent stops, and gives rise to a trajectory as follows:
%\begin{equation}\label{eq:traj}
%traj = (s_0, a_0, r_0, \cdots, s_t, a_t, r_t, \cdots),
%\end{equation}
%where the subscripts denote different timestamps over the finite time horizon.

\subsection{Problem Formulation}
In general, given a web application, the goal of testing is to generate action sequences along with suitable inputs in order to explore diverse states and behaviors of the application, potentially covering more logical application scenarios~\cite{FSE2019-WebTest}.

\begin{definition}[\textbf{Web State}]
A state $s$ is a description of a web application's current status (e.g., the HTML page).
\end{definition}

From a human perspective, the image (i.e., screenshot) that captures the changes of the HTML page is a natural representation of web states. However, due to the wide use of animations, images may not reliably represent web states (e.g., two completely different images may correspond to the same state of a web application). In comparison, the HTML page's code is a more precise representation as it encodes the URL and the structural characteristics of HTML pages. Thus, we propose a novel state representation by analyzing the HTML page's code (in Sec~\ref{sec:pre-procession}). Unless stated otherwise, a concrete HTML page instance is referred to as the state.

% Therefore, to better represent the state, we allocate a hash value to each HTML page by considering its code, which will be detailed in Sec~\ref{sec:pre-procession}. Unless stated otherwise, a concrete HTML page instance is referred to as the state, which has one hash value.
\begin{definition}[\textbf{Action}]
	An action $a$ is a valid operation in a given web state (i.e., an HTML page).
\end{definition}
\begin{figure}[t]
	\centering
	\includegraphics[width=.8\linewidth]{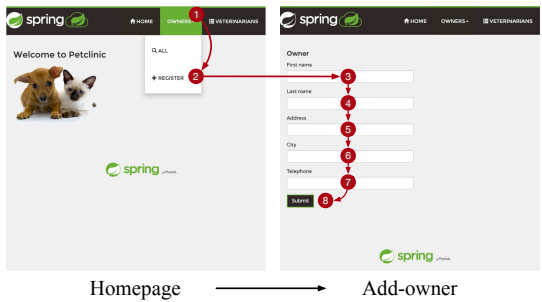}
	\caption{The action sequence of adding a new owner.}
	\label{fig:example}
\end{figure}

Given a web state, we focus on the operable DOM elements (e.g., links, buttons or input boxes), on which the operations may result in changes of application status (e.g., submitting a form or URL jumping). It is worth mentioning that, different states may contain different action DOM sets. For example, in Fig.~\ref{fig:example}, the homepage has 5 valid actions (i.e., elements in the navigation bar) while the ``add-owner'' page has 6 more valid actions (input boxes and a button).

\begin{definition}[\textbf{Test Case}] A test case is a sequence of actions $ (\mathit{a_0}, \cdots, \mathit{a_t}, \cdots)$ with necessary input values.
\end{definition}

For example, in Fig.~\ref{fig:example}, the ``add-owner'' function consists of three parts: navigating to the adding page, filling the form and clicking the submit button.
One feasible action sequence (in red) for testing the ``add-owner'' function is visualized in Fig.~\ref{fig:example}. This sequence together with necessary inputs constitute a test case $(a_0, \cdots, a_7)$. When operating on an inputtable element, our technique will provide a suitable value to make a test case continue, which will be detailed in the Section~\ref{sec:pre-procession}.

\label{sec:bug}
Similar to the existing work~\cite{FSE2019-WebTest}, we say a test case fails if exceptions or errors are thrown during the execution of the test case, including the JavaScript runtime exceptions, client errors or server errors, which can be analyzed via the returning status code~\cite{W3CErrorCode}. For simplicity, we refer to the exceptions and errors as failures in the subsequent sections.

\begin{definition}[\textbf{Web Testing}]
For a web application, the tester aims at learning an adaptive policy $\pi$ to continuously generate test cases for web exploration and failure detection.
\end{definition}

Intuitively, the more states are uncovered, the higher the chance of finding failures is.
Consequently, the goal of web testing is to generate test cases that can reach more diverse states, on which failures may be discovered.
It is worth emphasizing that discovering test cases that can trigger business logic (e.g., creating data) is critical for testing, as it may be a prerequisite for discovering new scenarios (e.g., editing data).

%% file: alg.tex
\section{Automatic Web Testing}
% In this section, we first give an overview of our approach. Then, we present each component and algorithm in detail.
\subsection{An Overview of \tool}
\begin{figure*}[t]
	\centering
	\includegraphics[width=\linewidth,height=4cm]{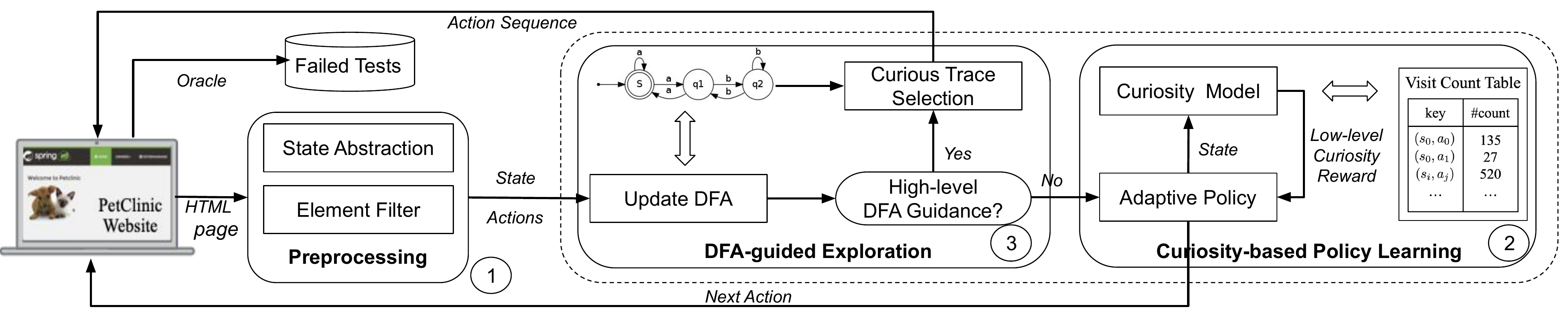}
	\caption{The Workflow of \tool.}
	\label{fig:overview}
\end{figure*}
In a nutshell, \tool is an end-to-end framework aiming at achieving automatic web testing in an online fashion. Its goal is to automatically generate diverse sequences of actions to explore more behaviors of the web application under testing. To achieve this, \tool leverage curiosity-driven reinforcement learning (RL) to constantly optimize a policy, which can generate diverse test cases. 
In particular, the RL training and web testing are intertwined, which is different from usual AI solutions that performs training before deploying.
Fig.~\ref{fig:overview} shows an overview of \tool, which comprises three major components. \circled{1} The \textit{pre-processing} component maps an HTML page to an abstract state. Its main purpose is to avoid the state explosion caused by dynamic updates in a web page, such that a good policy can be learned effectively. \circled{2} The \textit{curiosity-driven policy learning} component is designed for learning a policy that could explore diverse states of the web application. \circled{3} The \textit{DFA-guided exploration} component further improves the efficiency of the exploration of RL by maintaining a continuously updated deterministic finite automaton (DFA) that records all visited states and their frequencies. When RL cannot discover new states within a certain time budget, \tool selects one novel state as the starting point of the next exploration based on the global information of DFA, so as to avoid being trapped around the local optima.

Algorithm \ref{alg:apo} presents the details of our approach. \tool takes the target web application $env$ and the pre-processing function $\phi$ as the inputs and outputs a set of failed test cases $F$.
\tool first initializes the policy $\pi$, the DFA $M$, and the failed test set $F$ (Line~\ref{algo1:init}). Then, it continuously tests the web application until the pre-defined time budget exhausts (Line~\ref{algo1:timebudget}). During testing, to avoid reaching ``stuck webpages'' that cannot jump to other pages and continue further, we limit the maximum number of steps for one test case (Line~\ref{algo1:stepbudget}). After reaching the maximum number, we reset the web application: jump to the default homepage $p'$ (Line~\ref{algo1:reset}), set the initial action sequence, which includes only an empty action $\epsilon$ (Line~\ref{algo1:epsio}), and gets the initial state $s'$ (Line~\ref{algo1:initstate}). Each test case starts from the initial state, i.e., the homepage (Line~\ref{algo1:inititest}).
\begin{algorithm}[!t]
    \small
	\SetKwFor{Loop}{loop}{}{EndLoop}
	\SetKwInOut{Input}{Input}
	\SetKwInOut{Output}{Output}
	\SetKw{Continue}{continue}
	\SetKw{Break}{break}
	\Input{The target web application $env$, the pre-processing function $\phi(\cdot)$}
	\Output{The set of failed test cases $F$}

	Initialize policy $ \pi $, DFA \textit{M} and test case set $F=\emptyset$\\ \label{algo1:init}
	\Repeat{until time budget exhausts}{ \label{algo1:timebudget}
		$p': = reset(env)$\\\label{algo1:reset}
		$act := [\epsilon]$\\\label{algo1:epsio}
		$s':=\phi(p')$\\ \label{algo1:initstate}
		$t: = [s']$\\\label{algo1:inititest}
		\Repeat{reach maximum steps }{ \label{algo1:stepbudget}
			$p, \textit{failed}:= env(p’, act)$ \Comment{on-the-fly testing}\\ \label{algo1:testing}
			\If{failed }{
				$F:=F\bigcup\{t\}$\\	\label{algo1:addfailed}
			}

			%			sync q-table $ Q $ using state $ s $ and valid action set $ a_{set} $\\
			\circled{1} $s, va$ = $ \phi(p) $  \Comment{see Algorithm~\ref{alg:pre-processing}}\\   \label{algo1:validact}
			update valid actions of $ \pi(s) $ using $ va$\\
			\If{no curious state within some time}{
				\circled{3} $act:=selectTrace(M)$  \Comment{see Algorithm~\ref{alg:select-trace}}\\\label{algo1:selecttrace}
				$p'=reset(env)$\\\label{algo1:updatedefault}
				update the number of the current steps with the action sequence $act$\\\label{algo1:updatevalid}
				\Continue\\
			}

			\circled{2} calculate $ r = curiosity(s', a, s)$\\ \label{algo1:calcur}
			$a:=act[-1]$\\ \label{algo1:getaction}
			train policy $ \pi$ using $(s', a, r, s)$ \Comment{$Q$-learning}\\ \label{algo1:trainpolicy}

			update DFA \textit{d} using transition $ (s',a, s) $\\ \label{algo1:updatedfa}
			%			\Comment{$Q$-learning}
			$t.append(a,s)$ \Comment{store state-action sequence}\\ \label{algo1:updatetest}
			$act:= [\pi(s)]$ \\  \label{algo1:nextact}
			$ s' = s $\\ \label{algo1:curstate}
			$p'=p$\\ \label{algo1:curpage}
		}
		%	$C = C \cup \{c\}$ \Comment{Store test case to the set}\\
	}
	\Return $F$
	\caption{\tool}
	\label{alg:apo}
\end{algorithm}
 \tool performs on-the-fly testing by executing the current action sequence $act$ on the current page $p'$ (Line~\ref{algo1:testing}) (e.g., submitting a form). During the execution, \tool monitors the status of the browser's console such that failures can be captured automatically. %For example, in phoenix, the developer will log in console when there are some unexpected errors occur while socket connect happens.\yepang{Please explain phoenix. We should make the paper self-contained.}
 The environment returns a new HTML page $p$ and the error status. If an error is found, the test case is added into the failed test set (Line~\ref{algo1:addfailed}). \tool encodes the current page and returns the corresponding state $s$ and valid actions $va$ in the state $s'$ (Line~\ref{algo1:validact}). If \tool cannot identify a new state within a certain amount of time, the RL may enter into the local optima. \tool checks the DFA $d$ that records the global visit information, and selects one trace that is less visited (Line~\ref{algo1:selecttrace}). It returns the trace $t$ and the action sequence $act$, from which the trace can be restored. The web application is then reset to the homepage (Line~\ref{algo1:updatedefault}) and the number of the current steps is set with the length of trace $t$ (Line~\ref{algo1:updatevalid}).

After a state $s$ is explored, \tool calculates the curiosity reward $r$ (Line~\ref{algo1:calcur}). The policy $\pi$ is trained with the current transition $(s', a, s)$ and its reward $r$ (Line~\ref{algo1:trainpolicy}). In addition, the DFA $d$ and the current test case are updated with the transition (Lines~\ref{algo1:updatedfa}\textendash\ref{algo1:updatetest}). The next action is selected by feeding the current state $s$ to the policy $\pi$ (Line~\ref{algo1:nextact}). Then, the previous state and HTML page are updated (Lines~\ref{algo1:curstate}\textendash\ref{algo1:curpage}).

\subsection{Pre-Processing}\label{sec:pre-procession}
For policy learning via reinforcement learning, we need to define the state representation.
A straightforward way is to leverage web page representations (e.g., GUI or HTML document). However, if one adopts such a method, the number of states can be quite large and even infinite due to the dynamic nature of web applications.
For example, HTML documents can be different if the user operates differently (e.g., different form values or infinite scrolling pages).
Thus, adopting HTML document as states often suffers from the state explosion problem, resulting in low effectiveness of RL~\cite{bengio2013representation}. To overcome such a limitation, we propose a novel state representation. The intuition is that HTML pages that focus on the same business logic should be consolidated into one state. For example, in a webpage, the content of a table may be updated constantly (e.g., by adding or removing items). Although the HTML document may vary a lot, the pages still look similar and handle the same user interactions. We do not treat such pages as different states.
\begin{algorithm}[!t]
    \small
	\SetKwFor{Loop}{loop}{}{EndLoop}
	\SetKwInOut{Input}{Input}
	\SetKwInOut{Output}{Output}
	\SetKw{Continue}{continue}
		\SetKw{Return}{return}
	\Input{HTML page $p=\langle url, html\_doc\rangle$}
	\Output{State $s$, Valid action set $va$}
	Let $S$ be the current state set of the web application\\
%	$tag\_sequence := extract\_tags(html\_doc)$\\ %\Comment{Extract element tags from HTML}\\
	$va := retrieveValidElement(html\_doc)$\\ \label{algo2:retrive}

	\For{$s\in S$}{
		Fetch $\langle url', html\_doc’\rangle$ from the existing state $s$\\ \label{algo2:obtainstate}
		\If{$url \neq url'$}{\label{algo2:urlnoteql}
			\Continue \\
		}
	    $sim := computeSimlarity(html\_doc, html\_doc' )$\\ \label{algo2:sim}
	    \If{$sim > threshold$}{
	    	\Return $s, va$\\ \label{algo2:onestate}
    	}
	}
    Create new state $s$ using $\langle url, html\_doc\rangle$\\ \label{algo2:newstate}
    % $\gamma_s :=\langle url, html\_doc \rangle$ \\  \label{algo2:store}
    $S:=S\bigcup\{s\}$\\
	\Return $s, va$\\
	\caption{Pre-processing $\phi$}
	\label{alg:pre-processing}
\end{algorithm}

Given an HTML page, we use its \textit{URL} and \textit{HTML document} as the approximation of the business logic. It is intuitive that if two pages have the same URL and their HTML documents are very similar, they are more likely to focus on the same business logic. We argue that such similar pages represent the same state, and Algorithm~\ref{alg:pre-processing} describes how to distinguish different pages.
% Algorithm~\ref{alg:pre-processing} describes the pre-processing procedure including the state abstraction. 
The basic idea is to calculate the HTML structure similarity of two pages. If the structure similarity (via tag-wise comparison~\cite{GestaltPatternMatching}) is above a threshold, it is more likely that they focus on the same business logic and should be considered as the same state. 
% \yan{hash code?}
% \yan{Hence, webpages with identical URLs and similar HTML structure (via tag-wise comparison[32]) are treated as the same state, allocated with the identical hashcode.} 
Algorithm~\ref{alg:pre-processing} takes the code of the HTML page as the input and outputs the state $s$ as well as a set of valid actions $va$ in the current page. We use $S$ to represent the existing states during testing.

We adopt the browser built-in protocols~\cite{cdp} to filter some elements (e.g., no rendering or invisible) and only keep the ones that can be operated on, i.e., valid actions on this page (Line~\ref{algo2:retrive}). These elements include the clickable buttons, links, input boxes, selectors. Next, we check the similarity between the current page $p$ and pages in the previous states until one similar state is found. Specifically, for each previous state $s$, we obtain its page information (Line~\ref{algo2:obtainstate}) and calculate the similarity between $p$ and the page in $s$ as follows.
%(in our modeling, a state may correspond to many pages but we only save one page for each state for the efficiency of similarity comparison).
Firstly, we compare their URLs. Intuitively, if the URLs are not the same, the pages often tend to execute different business logic. Hence, we do not count them to be the same state (Line~\ref{algo2:urlnoteql}).
Otherwise, we calculate the similarity between the two HTML documents (Line~\ref{algo2:sim}). More specifically, we convert an HTML document to a sequence of tags and adopt the gestalt pattern matching~\cite{GestaltPatternMatching} algorithm to calculate the similarity between two sequences. If the similarity is above a pre-defined \textit{threshold}, the previously existing state $s$ and $va$ are returned (Line~\ref{algo2:onestate}). Note that, we extract all tags in the HTML document without any filtering so that no feature information in HTML will get lost.
If the current page does not match any existing states, we create a new state $s$ using the current page's code (Line~\ref{algo2:newstate}), add it into $S$ and return the results. 
It is worth noting that different URLs may represent the same business-logic, creating multiple states corresponding to the same business-logic (Line~\ref{algo2:urlnoteql}). However, such a correspondence causes little performance loss and doesn't affect \tool's soundness.
% and store the current page into the state (Line~\ref{algo2:store}). 
% Note that, each state, when created, is allocated with an unique hash code. 
% Note that for each state, our algorithm only stores the first HTML page belonging to the state instead of storing all similar pages.% Although we can store every page's information (i.e., HTML document) into a state, such a design will cause inefficiency: when checking whether a page belongs to an existing state, we need to perform many comparisons since a state may represent a large number of similar pa

Meanwhile, \tool focuses on generating action sequences rather than input values, even though input values can also affect the testing procedure.
To be aligned with prior works~\cite{FSE2019-WebTest,mesbah2011invariant} and carry out testing procedure, when operating on inputtable elements, random values will be generated according to the W3C standards~\cite{input-element}. Note that, \tool can leverage dictionaries or user-specified values to enhance the testing capability, which will be studied as the future work.

\subsection{Testing via Curiosity-Driven RL}\label{sec:alg}
\tool leverages RL to achieve an end-to-end testing by directly interacting with the web applications.
Specifically, the purpose is to learn a policy (i.e., $\pi$) that provides an exploration strategy to generate diverse test cases. To achieve such a policy, we need to define an effective reward function that determines the optimal policy.

\noindent\textbf{Reward function.} Common RL tasks (e.g., game playing) usually have a concrete goal such as winning a game or achieving a high score, which eases the design of the reward functions~\cite{YangHMZZZ19,ZhengHZMYLF21}. However, in web testing, reward function design becomes challenging as the goal is vague, i.e., to explore as many different behaviors of the web applications as possible. Moreover, a web application can be dynamically updated, indicating that the goal should also be dynamically adjusted. To address the challenge, we leverage the notion of curiosity, which has been proposed to counter the problem of coarse reward in RL~\cite{pathak2017curiosity,Savinov:2019tj}. Specifically, we have devised a curiosity-driven reward function that adopts a general and adaptive mechanism to guide the exploration such that diverse states could be reached. For curiosity measurement (Line~\ref{algo1:calcur} in Algorithm~\ref{alg:apo}), during testing, \tool maintains a visit count table to record the number of each transition (denoted by $ N(s',a, s) $).
The curiosity is measured by MBIE-EB~\cite{bellemare2016unifying}:
\begin{equation}\label{eq:curiosity}
curiosity(s',a, s) = \frac{1}{\sqrt{N(s', a, s)}}.
\end{equation}
$ N(s',a, s)$ is initialized to 1. Each time when the state $s'$ transits to $s$ by performing the action $ a $, the corresponding $ N(s',a, s) $ is increased by $1$.

\noindent\textbf{Q-Learning.} \tool leverages a model-free RL algorithm $Q$-learning~\cite{watkins1992q} to optimize the policy with the curiosity-driven reward. $Q$-learning has a function $Q:S\times A\rightarrow \mathbb{R}$, which returns the $Q$-value for a state-action pair.
Each time a new state $s$ is reached from the previous state $s'$ (i.e., $(s',a,s)$), we update the $Q$ function (Line~\ref{algo1:trainpolicy} in Algorithm~\ref{alg:apo}):
\begin{equation}\label{eq:q-learning}
Q(s',a) = curiosity(s',a) + \lambda max_{a^\prime} Q(s,a^\prime),
\end{equation}
where $ \lambda \in [0,1]$ is a discount factor. The $Q$ function keeps the temporal relations between actions since the $Q$-values will propagate to the ones in antecedent states recursively.

Based on the $Q$ function, the policy $\pi$ measures the weights of the valid actions in a state $s$ using the Gumbel-Softmax method~\cite{jang2017categorical}:
\begin{equation}\label{eq:gb-softmax}
p(a) = \frac{\exp(\frac{Q(s,a) + g_a}{\tau})}{\sum_{a_i\in \mathbf{A}}\exp(\frac{Q(s,a_i) + g_{i}}{\tau})}
\end{equation}
where $ \mathbf{A} $ is the valid action set at the state $ s $, $ \tau = 1 $ is a temperature coefficient, and $g_{(\cdot)}$ are i.i.d noise sampled from Gumbel$(0, 1)$ distribution. An action with a higher $Q$-value is more likely to be selected for interaction (Line~\ref{algo1:nextact} in Algorithm~\ref{alg:apo}).

This enables \tool to balance between exploration and exploitation. Actions that discover a new state are given a high curiosity reward and are more likely to be selected for the execution. This trait ensures efficient exploitation as such actions have high possibility in discovering diverse behaviors. In the meantime, the curiosity decreases along with the action execution, making other less executed actions to be selected for execution. This facilitates sufficient exploration and helps to find complex business logic in the web application, as we will demonstrate in our evaluation.

\begin{figure}[t]
	\centering
	\includegraphics[width=\linewidth]{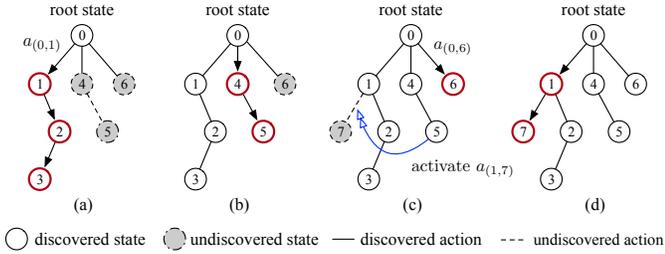}
	\caption{An intuitive illustration of curiosity model.}
	\label{fig:cm}
\end{figure}
\noindent\textbf{Example.} Fig.~\ref{fig:cm} illustrates how the curiosity-based RL works for web testing and how it can explore complex business logic.
Assume that \tool starts from the root state $s_0$ and different action $a\sim \pi(s)$ can be selected for execution. Executing actions cause state transitions, e.g., $ a_{(0,1)}$ results in $s_0 \rightarrow s_1$. Initially, the probability of choosing $a_{(0,1)}$, $a_{(0,4)}$, and $a_{(0,6)}$ are the same. Assume that $ s_3 $ is firstly covered through the red path (Fig.~\ref{fig:cm}(a)), then the value of $ Q(s_2, a_{(2,3)})$, $ Q(s_1, a_{(1,2)})$, and $ Q(s_0, a_{(0,1)})$ will be updated via back-propagation through the path. 
In this way, the temporal relations (along this path) is built and encoded in the policy $ \pi $. The curiosity reward $curiosity(s_0,a_{(0,1)},s_1)$ is decreased. In Fig.~\ref{fig:cm}(b) and Fig.~\ref{fig:cm}(c), actions $a_{(0,4)}$ and $a_{(0,6)}$ (with higher curiosity) are selected. Note that, after $s_6$ is reached, some previously invalid actions (e.g., $a_{(1,7)}$) may become valid due to the specific business logic. For example, an item in the table can only be deleted after being added. In this case, as $ curiosity(s_0,a_{(0,6)}, s_6) $ decreases, $ a_{(0,1)} $ and $ a_{(0,4)} $ regains the same chance to be selected, which is critical for exploring the newly activated states (i.e., $s_7$), opening a new untouched area to be explored.

\subsection{DFA-Guided Exploration}
\begin{figure}[!t]
	\centering
	\includegraphics[width=.9\linewidth]{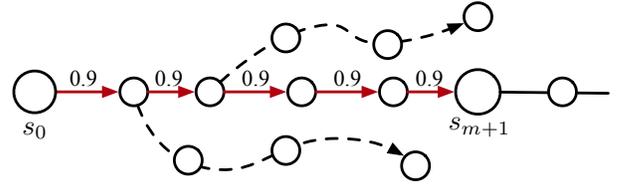}
	\caption{An intuitive illustration of transition in long path.}
	\label{fig:automation}
\end{figure}

Exploration is widely regarded as one of the most challenging problems of reinforcement learning~\cite{sutton2018reinforcement,pathak2017curiosity,Savinov:2019tj}. In the web application, a function is usually triggered by executing the actions in the specific order (e.g., the approval process in the Office Automation (OA) system). The exploration becomes more challenging as the sequence of actions gets longer. Although the curiosity-driven reward function provides guidance for the action selection, due to its stochastic nature, RL may still have low probability to select other actions especially in a long sequence of actions, interrupting the testing of the target function. Consider the example in Fig.~\ref{fig:automation}, a policy $ \pi $ with $0.9$ probability of selecting right actions (red arrow) to state $ s_{m+1} $. However, the possibility of reaching $ s_{m+1} $ is only $(0.9)^5=0.53$ since the path can be interrupted whenever the policy takes an action. The longer a path becomes, the more frequent an interruption may occur, making reaching a desirable transition harder, especially when facing a long path.

To address this challenge, we propose to build an on-the-fly deterministic finite automaton (DFA) during the testing, which provides a high-level guidance for boosting the RL exploration. Specifically, if new states cannot be found after some time, \tool starts to find a transition, which has the highest curiosity, from the DFA. Then, the shortest path that can reach the transition is detected such that the RL could directly reach this transition.

\begin{definition}[\textbf{DFA}] A deterministic finite automaton (DFA) $M$ is a 5-tuple $(S, A, \delta, s_0, F)$, where $S$ is a finite set of states, $A$ is a set of actions, $\delta: S \times A \rightarrow S$ is a set of transitions, $s_0$ is the initial state and $F$ is a finite set of states that cannot transit to other states.
\end{definition}

\begin{algorithm}[!t]
    \small
	\SetKwFor{Loop}{loop}{}{EndLoop}
	\SetKwInOut{Input}{Input}
	\SetKwInOut{Output}{Output}
	\SetKw{Continue}{continue}
	\SetKw{Return}{return}
	\Input{DFA $ M $}
	\Output{Action set $act$}
% 	$C: = {curiosity(s) | \forall (s', a, s)\in \delta}$\;
	$(s_m,a_m,s_{m+1}) := \argmax_{(s',a,s)\in \delta} (curiosity((s',a,s)))$\;
	Find the shortest trace $tr:= (s_0,a_0,s_1,a_1,\ldots,s_m,a_m,s_{m+1})$\;

	\Return $(a_0, a_1,\ldots,a_m)$\;
	\caption{selectTrace}
	\label{alg:select-trace}
\end{algorithm}
During testing, once a new transition $(s',a, s)$ is explored, the DFA will be updated, i.e., $\delta:=\delta \bigcup \{(s',a, s)\}$ (Line~\ref{algo1:updatedfa} of Algorithm~\ref{alg:apo}). Algorithm~\ref{alg:select-trace} presents the basic idea of curiosity-driven trace selection. From the DFA, we first select the transition with the highest curiosity $(s_m, a_m, s_{m+1})$. Then, we adopt the Dijkstra's algorithm~\cite{bondy1976graph} to identify the shortest trace $tr$ that can reach the target $(s_m, a_m, s_{m+1})$. With this trace, RL could directly restore to the target transition. Intuitively, some transitions 
could be very deep and thus difficult to reach by RL. For these transitions, \tool leverages DFA to further enhance the exploration efficiency and testing effectiveness.

Theoretically, the non-deterministic finite automaton can represent the dynamicity of the stochastic environment more accurately. However, we use DFA due to the following reasons: 1) the automaton is used to guide the selection of one viable path for exploration. Due to dynamic factors (e.g., network), one path in DFA may be infeasible, but it doesn't affect the soundness of \tool because the dynamic execution will ignore such infeasible paths; 2) The construction of non-deterministic finite automaton could be more expensive, especially on estimating the transition probabilities. Considering the trade-off between efficiency and granularity of automaton construction, DFA is a good-enough solution for \tool.

%% file: evaluation.tex
\section{Empirical Evaluation}
We have implemented \tool based on  Python 3.7.6 and PyTorch 1.5.0~\cite{paszke2017automatic} with more than 5,000 lines of code\footnote{More details can be found in our website \cite{DeepExplorer}.}.
To demonstrate the effectiveness and efficiency of \tool, we conduct an empirical evaluation investigating the following four research questions.

\begin{itemize}[leftmargin=*]
\item \textbf{RQ1 (Code Coverage):} How is the exploration capability of \tool in terms of code coverage?

\item \textbf{RQ2 (Failure Detection):} How effective is \tool for detecting failures of web applications?

\item \textbf{RQ3 (DFA Guidance):} How effective is DFA in guiding the exploration during testing?

\item \textbf{RQ4 (Scalability):} How effective is \tool in testing real-world web applications?
\end{itemize}

\subsection{Experiment Setup}
\subsubsection{Benchmarks}
Our large-scale evaluation uses three benchmarks, including a research benchmark from the prior work~\cite{FSE2019-WebTest} to compare \tool with the state-of-the-art techniques, a benchmark of top 50 real-world web applications~\cite{aleax} to evaluate the scalability of \tool, and an industrial web application to conduct a detailed case study.

\begin{itemize}[leftmargin=*]
\item \textbf{Research benchmark:} We adopt a benchmark containing six popular GitHub projects (each has more than 50 stars) from the prior work~\cite{FSE2019-WebTest}. These projects use six most popular JavaScript frameworks: dimeshift (Backbone.js), pagekit (Vue.js), Splittypie (Ember.js), phoenix-trello (Phoenix/React), Retroboard (React), and PetClinic (AngularJS).

\item \textbf{Real-world web applications:} According to the ranking~\cite{aleax}, we select the top 50 web applications in the world for evaluation. To investigate the scalability, we directly leverage \tool for an end-to-end testing of these applications without fine-tuning.

\item \textbf{Industrial web application:} A complex Software as a Service (SaaS) system is adopted for the further case studies. We omit the system name for anonymous review reasons.
\end{itemize}

\subsubsection{Web application failures}\label{sec:fault-definition}
In subsequent experiments, we collect the system-level failures (defined in Section~\ref{sec:bug}) reported in the browser's console to study the failure detection capability of related approaches.
Note that user-level failures may or may not cause system-level failures, which depends on the system's robustness. For example, if user-level failures are well handled by the web-system (e.g., strict input-field checking), no system-level failures will occur. Otherwise, failures will be triggered and captured by \tool. For identifying the root-cause of failures (e.g., by users or system), we adopt manual analysis.
It is worth emphasizing that all discovered failures are manually vetted to ensure that the thrown exceptions and errors are actually failures (i.e., no false alarms).

\subsubsection{Baselines Approaches}
To evaluate the effectiveness of \tool, we select three state-of-the-art approaches as baselines for a comparative study. These baselines include both the model-based to model-free algorithms. Besides, one random strategy that adapts the idea of Monkey~\cite{monkey} is adopted as a baseline.
Moreover, to evaluate the advantage of leveraging DFA, a variant of \tool is also implemented for the ablation evaluation.
\begin{itemize}[leftmargin=*]
	\item DIG~\cite{FSE2019-WebTest} is a navigation model-based approach, leveraging a diversity-based test case generator for web testing.
	\item SUBWEB~\cite{biagiola2017search} is a navigation model-based approach, considering the uncovered branches and using a search-based strategy to achieve web testing.
	\item Crawljax~\cite{crawljax} is a navigation model-free approach, discovering and clustering pages on the fly and adopting a crawling-based random test case generator for web testing.
	\item Random~\cite{monkey} is a model-free approach, randomly selecting one of the available actions to explore web states.
	\item \tool(no DFA) is a variant of \tool without the DFA guidance.
\end{itemize}

\subsubsection{Configurations}\label{exp:configuration}
For all experiments, we give each tool the same time budget (i.e., 30 minutes). To counteract the randomness from a statistical perspective, we repeat each experiment 15 times and calculate the average results. For the similarity in the pre-processing, we set 0.8 as the threshold. DFA provides high-level guidance for \tool if no new states are discovered in 2 minutes. For the discount factor in RL, we set $\lambda$ = 0.95 in all experiments.
We conduct an comprehensive evaluation in spending more than 300 CPU hours, i.e., 6 projects * 7 settings (for 5 tools) * 0.5 (hour time budget for single round) * 15 repetitions in total. Besides, during testing, actions that lead to external links (via domain checking) will be recorded during testing, marked as invalid actions, and not executed in the subsequent testing.

\subsection{Code Coverage (RQ1)}\label{sec:RQ1}
To conduct a comprehensive comparison, for DIG and SUBWEB, we use the naviation models, which are
based on automatically and manually generated page objects (denoted by APO and MPO), respectively.
To counteract implementation bias, both the APO and MPO are directly adopted from the prior work~\cite{FSE2019-WebTest}.
Comparisons in terms of the branch coverage of JavaScript code are conducted on six web applications.
The averaged results of 15 runs are summarized in Table~\ref{tab:coverage}, where bold numbers indicate the best result.
Overall, we have the following findings.
\def\arraystretch{1.2}
\begin{table*}[t]
    \fontsize{5.8pt}{5.8pt}\selectfont
	\begin{center}
		\caption{Comparisons of related baselines regarding the averaged branch coverage and failure detection with the corresponding standard deviation.
		(Values in bold indicate the best average results using 15 runs.)
		}
		\label{tab:coverage}
        \centering
        \renewcommand{\arraystretch}{1.6}
        \begin{tabular}{c||ccccccc||ccccccc}
            \hline 
            \multicolumn{1}{c||}{\multirow{3}{*}{Subjects}} & \multicolumn{7}{c||}{Average Branch Coverage (\%)} & \multicolumn{7}{c}{Average Unique Failures (\#)}\tabularnewline
            \cline{2-15} \cline{3-15} \cline{4-15} \cline{5-15} \cline{6-15} \cline{7-15} \cline{8-15} \cline{9-15} \cline{10-15} \cline{11-15} \cline{12-15} \cline{13-15} \cline{14-15} \cline{15-15} 
             & \tool  & Crawljax  & Random  & \multicolumn{2}{c}{DIG} & \multicolumn{2}{c||}{SUBWEB} & \tool  & Crawljax  & Random  & \multicolumn{2}{c}{DIG} & \multicolumn{2}{c}{SUBWEB}\tabularnewline
            \cline{2-15} \cline{3-15} \cline{4-15} \cline{5-15} \cline{6-15} \cline{7-15} \cline{8-15} \cline{9-15} \cline{10-15} \cline{11-15} \cline{12-15} \cline{13-15} \cline{14-15} \cline{15-15} 
             & \multicolumn{3}{c}{Navigation model-free} & \multicolumn{1}{c}{APO } & MPO  & APO  & MPO  & \multicolumn{3}{c}{Navigation model-free} & \multicolumn{1}{c}{APO } & MPO  & APO  & MPO\tabularnewline
            \hline 
           Dimeshift  & \textbf{51.0} (2.7)  & 11.8 (0.0)  & 24.1 (12.8)  & 38.5 (1.6)  & 40.1 (1.6)  & 36.7 (2.3)  & 38.9 (1.8)  & \textbf{9.0} (1.7)  & 1.0(0.0)  & 1.4 (0.5)  & 2.6 (0.8) & 2.9 (0.5)  & 2.4 (0.9)  & 2.9 (0.5) \tabularnewline

            Pagekit  & \textbf{36.0} (1.3)  & 0.5 (0.0)  & 0.5 (0.0)  & 24.3 (2.6)  & 31.7 (4.9)  & 24.7 (2.8)  & 27.6 (3.8)  & \textbf{4.2} (0.9)  & 0.0 (0.0)  & 0.0 (0.0)  & 1.0 (0.0)  & 2.9 (0.4)  & 1.0 (0.0)  & 0.7 (1.0) \tabularnewline

            Splittypie  & 43.4 (0.3)  & 37.7 (3.9)  & 18.9 (1.1)  & \textbf{46.9} (1.8)  & 45.3 (2.5) & 44.1 (3.5)  & 44.7 (2.0)  & \textbf{8.0} (1.4)  & 5.6 (0.5)  & 3.0 (0.0)  & 4.0 (0.0)  & 4.0 (0.0)  & 4.0 (0.0)  & 6.0 (0.0) \tabularnewline

            Phoenix  & \textbf{81.7} (1.3)  & 30.2 (9.2)  & 42.1 (0.0)  & 60.1 (5.7)  & 65.1 (9.6)  & 61.3 (2.3)  & 63.1 (8.7)  & \textbf{5.3} (0.5)  & 0.0 (0.0) & 1.2 (0.0)  & 2.0 (0.0)  & 2.0 (0.0) & 2.0 (0.0)  & 2.1 (0.5) \tabularnewline
            
            Retroboard  & 61.4 (0.3)  & 22.2 (0.0)  & 56.1 (0.0)  & 68.7 (2.1)  & 70.9 (4.3)  & 68.1 (2.7) & \textbf{71.9} (3.8)  & \textbf{1.0} (0.0)  & 0.0 (0.0) & \textbf{1.0} (0.0)  & \textbf{1.0} (0.0)  & \textbf{1.0} (0.0)  & \textbf{1.0} (0.0)  & \textbf{1.0} (0.0) \tabularnewline
            
            Petclinic  & \textbf{85.0} (0.0)  & 18.0 (2.5)  & 35.0 (26.3)  & 80.5 (7.3)  & 49.5 (7.9) & 83.0 (2.9)  & 49.5 (7.9)  & \textbf{11.7} (1.0)  & 0.7 (0.0)  & 2.0 (0.0)  & 1.5 (0.0)  & 3.0 (0.7)  & 1.5 (2.1)  & 3.3 (0.7) \tabularnewline
            
            \hline 
            Average  & 59.8  & 20.1  & 29.45  & 53.2  & 50.4  & 53.0  & 49.3  & 6.5  & 1.2  & 1.4  & 2.0  & 2.6  & 2.0  & 2.7 \tabularnewline
            \hline 
            \end{tabular}

	\end{center}
\end{table*}

The model-based methods DIG and SUBWEB can achieve an overall better branch coverage than model-free methods Crawljax and Random in most cases. This is because navigation models can provide more information (e.g., web structure), which is beneficial for effective testing. However, a counterfactual finding is that model-free \tool achieves competitive performance in terms of code coverage, significantly outperforming  (i.e., calculated by Mann-Whitney U test~\cite{u-test} at 0.05 confidence level) model-based methods in 4/6 web applications (bold numbers in Table~\ref{tab:coverage}).
It not only demonstrates the exploration capability of \tool in terms of the code coverage, but also the robustness, which could be obtained by the model-free testing fashion.

Besides, we perform an in-depth analysis on the test cases generated by \tool to figure out why \tool, as a model-free algorithm, can achieve better performance than other model-free algorithms (i.e., Crawljax and Random). Take Petclinic for an example (shown in Fig.~\ref{fig:example}), we find that \tool can generate the correct operation sequence (i.e., filling form values before adding an owner) to achieve effective testing.
Normally, generating such logically related sequence actions is hard for random-based model-free algorithms.
However, by leveraging the curiosity-driven RL, \tool can capture such relations and encode this ``knowledge'' in the policy to create effective test cases without navigation models.

Furthermore, compared to model-based algorithms (i.e., DIG and SUBWEB), we found that \tool performs much better in four subjects and similarly for the rest.
We investigate the reason and find that some pages in Splittypie and Retroboard need complex inputs that are difficult to generate randomly. For instance, the ``transaction'' page in Splittypie can only be discovered after typing in a time value with a \textit{non-standard} W3C format (e.g., mmdd). However, \tool follows W3C standards~\cite{input-element} and cannot generate such inputs without human knowledge. Meantime, we analyze APO and MPO in DIG and SUBWEB, and find that both kinds of navigation models have been manually fine-tuned with human-knowledge, which enables non-standard input generation for higher coverage. Detailed results and analysis can be found in \cite{DeepExplorer}. In rest four subjects, such non-standard inputs barely exist, where \tool achieves much better results.
\begin{tcolorbox}[size=fbox,arc=0mm,outer arc=0mm,boxrule=.3mm]
	{\textbf{Answer to RQ1:} In contrast to model-based approaches, \tool achieves better code coverage and robustness in most cases with no navigation models or prior knowledge.
	}
\end{tcolorbox}

\subsection{Failure Detection (RQ2)}
We continue to analyze the ability of each baseline in discovering failures (defined in Section~\ref{sec:fault-definition}). Table~\ref{tab:coverage} shows the statistical results of the average number of failures discovered during the allocated testing time (see Section~\ref{exp:configuration}).
As one failure can be discovered multiple times during the entire testing process, here we only count the unique failures for all methods.
Overall, we have the following findings.

First, compared with other baselines, random approach achieves relatively poor performance, which is consistent with the intuition that simple random exploration is ineffective for large web applications.
Model-based approaches can discover more failures than the random approach but exhibit instability.

Among all baselines, \tool discovers the most number of failures (bold numbers in Table~\ref{tab:coverage}), significantly exceeding other methods (i.e., calculated by Mann-Whitney U test at 0.05 confidence level). This reveals not only the competitive performance in failure detection, but also generality of \tool.

Another counterintuitive finding is that, in Splittypie and Retroboard subjects, although \tool achieves lower code coverage than DIG (in Table~\ref{tab:coverage}), it still detects more failures (in Table~\ref{tab:coverage}).
%\tool achieves better failure detection rate (in Table~\ref{tab:bug-detection}) but lower code coverage than DIG (in Table~\ref{tab:coverage}).
We analyze the test cases generated by \tool and DIG, compare the logs (see detailed logs in~\cite{DeepExplorer}) and find that \tool discovers more server request errors (e.g., error code $400$ and $500$) via generating test cases with illegal operation sequences. Intuitively, the code related to each operation can be easily covered by independent execution. However, some failures are only triggered by illegal operation orders, indicating that it seems ineffective to detect failures by simply improving code coverage.
Moreover, DIG combines a sequence of actions with a specific order as a macro operation in both APO and MPO, ignoring the fact that different orders or executing only part of the sequence may result in potential failures. This explains why \tool performs better in generating effective test cases.
\begin{figure*}[t]
	\centering
	\includegraphics[width=\linewidth]{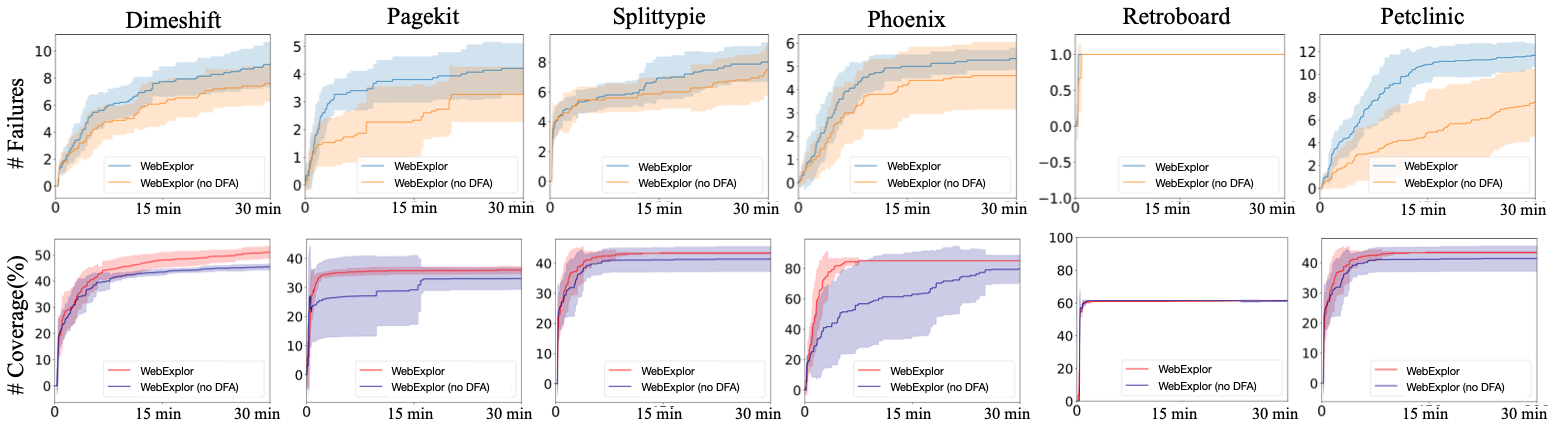}
	\caption{Evaluation of DFA regarding averaged number of discovered failures (top), code coverage (bottom) and testing efficiency (15 runs). The results are averaged using 15 runs, while the solid line and shaded represent the mean value and standard deviation, respectively. }
	\label{fig:dfa-guidence}
\end{figure*}

\begin{tcolorbox}[size=fbox,arc=0mm,outer arc=0mm,boxrule=.3mm]
	{\textbf{Answer to RQ2:} Compared to other baselines, \tool achieves the best performance in failure detection, while requiring no human knowledge or fine-tuned navigation models, revealing its potentials across different subjects.}
\end{tcolorbox}

\subsection{DFA Guidance (RQ3)}
This section investigates how DFA contributes to boosting the testing efficiency through high-level guidance.
Comparisons between \tool and \tool(no DFA) are conducted to evaluate the failure detection rate and efficiency. Fig~\ref{fig:dfa-guidence} illustrates the results, where the x-axis is the testing time and y-axis (\#Failures and \#Coverage) are the average number of discovered failures and code coverage rate, respectively.
To avoid statistical bias, the results are averaged using 15 runs, and the bold lines and shadow areas represent the mean and standard deviation.

First, in Fig.~\ref{fig:dfa-guidence} (top), we observe that \tool can not only discover more failures than the one without DFA, but also a higher failure detection efficiency (blue line rises faster). Meanwhile, DFA achieves an early performance jump regarding the number of discovered failures, especially in the dimeshift, pagekit and petclinic subjects. All the advantages benefit from a better exploration guided by DFA.
Take the petclinic subject for example, many failures discovered by \tool have long execution traces, containing many actions that need to be executed in a specific order. Consider this page transition trace as an example: homepage $\rightarrow$ owner-list $ \rightarrow $ owner-info $ \rightarrow $ pet-list $ \rightarrow $ pet-info $ \rightarrow $ visit-info page. Many failures can only be discovered when entering the visit-info page and submit a form with invalid values. \tool can efficiently achieve this transition via DFA guidance and start subsequent testing. Otherwise, any interruption in the process will result in visiting another page, degrading the testing efficiency (e.g., Fig.~\ref{fig:cm}). Therefore, such high-level guidance effectively leads to promising testing directions, by which \tool achieves efficient and effective web testing.

%Similar improvement regarding the code coverage can be observed in Fig. ~\ref{fig:dfa-guidence} (bottom).
Another finding is that DFA can achieve a more stable performance in terms of both failure discovery and code coverage. As shown in Fig. \ref{fig:dfa-guidence}, the standard deviation (shaded area) of using DFA is smaller than not using DFA. The variation is particularly obvious in pagekit, phoenix and petclinic subjects.

%In spite of detecting more bugs, \tool also achieves a faster bug detection speed, where \tool in blue line achieves an early jump especially in the dimeshift, pagekit and petclinic subjects, indicating that DFA guidance indeed boost the testing efficiency of \tool.
\begin{tcolorbox}[size=fbox,arc=0mm,outer arc=0mm,boxrule=.3mm]
	{\textbf{Answer to RQ3:} DFA complements curiosity-driven RL by providing high-level guidance for effective exploration and high efficiency. \tool achieves a better performance in terms of failure detection rate, code coverage and stability.}
\end{tcolorbox}

\subsection{Evaluation on Real-World Web Applications (RQ4)}
RQ4 aims to evaluate the effectiveness of \tool on real-world web applications. According to the \textit{Alexa} rank list~\cite{aleax}, top 50 most popular web applications are chosen for evaluation.
In total, \tool discovered 3,466 failures.
%which are shown in Table~\ref{tb:rq3-summary}.
%We collect the bugs (defined in Section~\ref{sec:fault-definition}) discovered during online testing using \tool, which discovered 3,290 faults in total.
After manual inspection, we find these failures consist of 1,889 JavaScript failures, 147 server failures (i.e., status code $\geq$ 500) and 1,430 other failures (more details in \cite{DeepExplorer}).
Moreover, we analyze the URLs of the detected failures and find that 83.71\% failures come from the original web applications, while the rest are from third-party libraries. This indicates that most failures indeed exist in the original web applications, resulting in an urgent need of end-to-end testing such as \tool.
%an efficient bug detection tool like \tool.

On the other hand, we find that the failures can be caused by either the client or server sides, including static resources loading errors, JavaScript errors due to uncommon operation sequences, and crossing site errors due to accessing different servers (refer to~\cite{DeepExplorer} for details), which demonstrates the effectiveness of \tool in detecting failures in real-world web applications. Moreover, \tool leverages no manually-tuned parameters, demonstrating high scalability among various modern web applications.
More analysis and failure screenshots including web pages and console outputs can be found on our website~\cite{DeepExplorer}.

%\item {\textit{ There are 33.04\% bugs which are caused by the static resources loading (e.g., stylesheet, javascript, images).}} It indicates that static resources should be well maintained by developers.
%
%\item \textit{\textit{ There are 6.91\% bugs caused by accessing crossing site resources or API.}} It is very common especially for the cross-site request.
%
%\item \textit{\tool is very effective in detecting the JavaScript errors (19.85\%) by exploring more uncommon operation sequence.}
%\end{itemize}
%These findings prove the effectiveness of \tool in fault detection. Moreover, \tool uses no manual-tunes parameters, demonstrating a high scalability among modern web applications.

Furthermore, a commercial SaaS platform (over a million lines of code) is employed as a detailed case study. %for evaluation.
\tool successfully finds 12 unspotted failures, which are confirmed and fixed by developers. We analyze some specific cases of discovered failures as follows.

\noindent\textbf{Asynchronous rendering:} The following code segment shows a discovered failure, where gray lines are newly added fixing solutions. The elements (i.e., ``kg-container") is asynchronously rendered, which will be a null pointer before finishing rendering. Therefore, a failure will be triggered once accessing such a null pointer (Line 3). This finding also suggests that developers should pay attention when using asynchronous techniques in web applications.
\begin{lstlisting}[language=C,escapeinside=\`\`]
     // Disable default right click behavior
  `\colorbox{lightgray}{\rlap{+   if (!!document.getElementById("kg-container"))}\hspace{\linewidth}\hspace{-4\fboxsep}}
  \\`  // throws null point exception without checking
          document.getElementById("kg-container")
					.oncontextmenu = function(e) {
							e.preventDefault();
                };
\end{lstlisting}

\begin{lstlisting}[language=C,escapeinside=\`\`]
     // Check invalid email
  `\colorbox{lightgray}{\rlap{
    +   if (match\_email("kg-container"))}\hspace{\linewidth}\hspace{-4\fboxsep}}
    \\
        register\_email(email);
\end{lstlisting}

\noindent\textbf{Security defect:} The following example shows a module loading failure that results in a security defect. Specifically, when \tool operates on the ``online-editor'' in the browser, the server tries to load a non-existent module ``brace/mode/c\_cpp'' by traversing all folders (Line 3).

\begin{figure}[h]
	\centering
	\includegraphics[width=\linewidth]{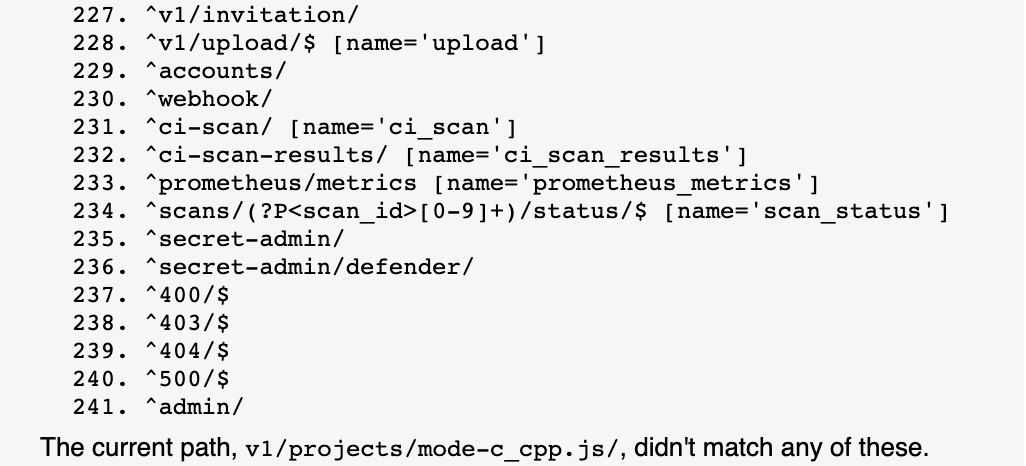}
	\caption{The exposure of sensitive server information.}
	\label{fig:security-defect}
\end{figure}

Consequently, as shown in Fig.~\ref{fig:security-defect}, the server exposes all folders and folder structure to the client, which could be exploited by malicious attackers, and results in security problems. The ``online-editor'' is difficult to reach but \tool discovered it by adopting an effective exploration.

%\yi{On the other hand, a security defect is spotted by \tool (As shown in Listing~\ref{lst:unavialable-module}).When \tool tries to access an online-editor, the unavailable module is loaded and redirected to a server with debug mode which exposing all of its directory.}

Beyond these, \tool discovered some other failures. For instance, a password-reset API throws an internal exception (400 bad request) once receiving an invalid email input.
Besides, when \tool executes the tab-switching behaviors, an event handler error is thrown, which results in incorrect data rendering.
Moreover, searching API throws an error (500 internal server error) if the request payload contains no product or organization ID, which cannot be handled. More analysis and details can be found in our website~\cite{DeepExplorer}. %These discovered and confirmed bugs further demonstrate the usefulness of \tool in testing real-world commercial web applications.

\begin{tcolorbox}[size=fbox,arc=0mm,outer arc=0mm,boxrule=.3mm]
	{\textbf{Answer to RQ4}: \tool tests modern web applications in an end-to-end fashion with no additional manual efforts (e.g., building a navigation model).
	Besides, 3,000+ failures detected from the real-world applications further demonstrate the effectiveness of \tool in failure detection.
	}
\end{tcolorbox}

\subsection{Threats To Validity}
Randomness can be a major threat during testing and the web interaction. We reduce this threat by repeating 15 times for each testing configuration and average the results.
Besides, the oracle we proposed may not be complete, and thus \tool may miss some other unknown failures like UI bugs or other types of bugs. 
As \tool performs black-box web testing, the server-side codes are assumed to be unavailable. Therefore, only the coverage of the client-side codes (e.g., JavaScript) are reported, and the evaluation of \tool in terms of the code coverage may be incomprehensive.
The selection of the RL algorithm used for training policies could be biased. We mitigate this problem by selecting the standard RL algorithm~\cite{watkins1992q} for web testing.
Furthermore, hyperparameters may be a potential threat. The choice of such hyperparameters is dependent on the domain knowledge, and may be biased. Besides, the selection of the web applications could be biased. To address this, a research benchmark and various real-world web applications, including active commercial websites, are adopted for evaluation.

%To address this, for each discovered failure, \tool records the necessary information (including every taken action, all encountered web pages during the whole testing, detailed information of network request, etc.).

%Moreover, an additional replay suit is also developed for replaying the web interaction automatically. On this basis, despite being unable to achieve 100\% detection accuracy, \tool provides an effective way for testers to discern real failures efficiently.

%% file: study.tex
\section{Related Work}
Many techniques have been proposed to automate web testing.
In the following, we briefly discuss the most relevant solutions and their limitations, which motivates the need for a novel and efficient web testing technique.

\noindent\emph{\textbf{Model-Based Testing.}}
Model-based technique is a major paradigm for achieving automatic web testing~\cite{FSE2019-WebTest,biagiola2017search,mesbah2011invariant,yu2015incremental}.
This kind of techniques build models to describe the web applications’ behaviors in advance, and then derive test cases from the models to find bugs. For instance,
%This kind of technique is a major paradigm for achieving automatic web testing, since the test cases can be generated underlying the constructed model without directly interacting with the web applications.
approaches like DIG~\cite{FSE2019-WebTest}, SubWeb~\cite{biagiola2017search}, and ATUSA~\cite{mesbah2011invariant} extract paths from the navigation model, where genetic algorithm is adopted for performing path selection and input generation. Further, an incremental two-steps algorithm InwertGen~\cite{yu2015incremental} is proposed, where the generation of navigation model and test cases are intertwined.
Overall, model-based techniques exhibit the advantage of fast test cases generation since no web interaction is required. However, the navigation model may cover only some behaviors of the web applications while others cannot be tested. Besides, expert domain knowledge are required in building high quality models.
%Hence the accuracy and completeness of the constructed model will be of great importance.
%\yan{Test cases are generated using the tool Randoop.}
Such limitations motivate the need for a model-free testing technique. \tool, to the best of our knowledge, is the first RL-based technique that performs an end-to-end automated web testing for real-world applications and achieves competitive performance comparing to the state-of-the-art techniques.

\noindent\emph{\textbf{Test Case Generation}}.  Test case generation consists of building test path and corresponding input values~\cite{FSE2019-WebTest}.
Given a navigation model, test paths can be generated by search-based approaches~\cite{biagiola2017search,dincturk2014model}, and inputs can be generated using random or evolutionary algorithms~\cite{FSE2019-WebTest,artzi2011framework}.
In many cases, search-based techniques need to explicitly address path feasibility, resulting in tremendous executions of test case candidates. Search-based algorithms also need to evaluate each test case's fitness value, which is costly since plenty of candidates need to be generated and executed in the browser before converging~\cite{luke2013essentials}.
%However, in modern web applications, the graph visit technique is a simple ways of generating test paths from a navigation model, which may not be able capture relations between complex pages. Both
%After that, random inputs together with the test path makes up a concrete test case for web testing~\cite{mesbah2011invariant}.
%However, search-based techniques iteratively sample the input space, selecting the fittest candidate test cases, and evolving the fittest ones using genetic search operators to create new test cases~\cite{luke2013essentials}.
%However, the evaluation of the fitness value is costly, since it requires numbers of candidates to be generated and executed in the browser before converging.
Existing techniques for automated test case generation either ignore path feasibility or require a large number of executions.
As for input values, random strategy~\cite{mesbah2011invariant,artzi2011framework} can generate feasible inputs in most cases.
Furthermore, symbolic execution techniques (e.g., Apollo~\cite{artzi2008finding}, Jalangi~\cite{sen2013jalangi} and SymJS~\cite{li2014symjs}) use systematic strategies to generate specific inputs to cover hard-to-reach codes.
However, generating such specific inputs is not the primary concern of \tool, and all advanced input generation techniques can be incorporated in the \tool to further enhance the performance. Recently, there are also some research on testing deep learning models~\cite{DeepExplorer,deephunter19,deepstellar,xie2019diffchaser}, which could be used to test reinforcement learning models.

%Randomly generated inputs have often low chance of producing feasible test cases~\cite{mesbah2011invariant,artzi2011framework}. Therefore, a huge number of input generations and corresponding test executions are required~\cite{biagiola2017search,marchetto2008state}. However, \tool focuses on generating effective test path. The corresponding input values are not the primary concern.

\noindent\emph{\textbf{Reinforcement Learning Based Testing}}.
In recent years, RL has also been applied to help software testing. For instance, Wuji~\cite{zheng2019wuji} leverages RL and multi-objective optimization for game testing, which is rather different from web applications.
B{\"{o}}ttinger et al.~\cite{BottingerGS18} introduce the first RL based fuzzer to learn high reward seed mutations for testing traditional software.
Retecs~\cite{SpiekerGMM17} uses RL to facilitate test prioritization and selection in regression testing.
Besides, RL is also adopted in mobile testing techniques~\cite{KorogluSMMUTD18,pan2020reinforcement,AdamoKKB18,VuongT18}, ranging from QBE~\cite{KorogluSMMUTD18} for crash detection to GUI testing~\cite{AdamoKKB18,VuongT18}. The most recent Q-testing~\cite{pan2020reinforcement} achieves the best mobile testing performance by leveraging the curiosity reward, which is similar to \tool. Both \tool and $Q$-testing are built on the $Q$-learning, however, due to different characteristics between Android and Web applications, the challenges addressed by the two works are different. In \tool, we carefully design state representation and suitable rewards for web testing, which are different from Q-testing. 
Furthermore, we also found that it is not effective when only using $Q$-learning. 
Thus DFA-augmented exploration is proposed in \tool, which is also a key difference compared with $Q$-testing.
The existing studies most related to both the RL and web domains are Artemis~\cite{liu2018reinforcement} and DOM-Q-NET~\cite{jia2019dom} that utilize RL to perform explicitly defined web-based tasks.
Different from completing explicit tasks, \tool targets at web testing via exploring diverse behaviors using curiosity-driven RL, which has not been tackled before.

% However, Q-testing uses UIAutomator~\cite{uiautomator} to extract GUI hierarchy, which can not be directly applicable in web testing due to the difference between domains.
% In addition, \tool adopts DFA to guide effective and efficient web testing and exploration.

% Both WebExplor and Q-testing adopt reinforcement-learning for software-testing. The common part is the use of Q-learning, which is an existing technique (not the contribution of these two works). 
% However, due to different characteristics between Android and Web applications, the challenges addressed by the two works are different. 
% In particular, we need to carefully design the state representation and suitable rewards for web testing, which are different from Q-testing. 
% We also tried Q-testing and found it hard to adapt it to the Web-domain. 

%proposes the curiosity-driven RL with the automation guidance to improve testing of web applications, which has not been tackled before.